\documentclass[a4paper,onecolumn,showpacs,amsmath, amssymb,nofootinbib,aps,floatfix]{revtex4}
\usepackage{amsmath}
\usepackage{amsfonts}
\usepackage{amssymb}
\usepackage[dvips]{graphicx}
\usepackage[dvips]{color}
\usepackage{epsfig}
\input psfig.sty 

\usepackage{bm}                                 
\usepackage{dcolumn}                                
\usepackage{hyperref}

\newcommand{\no}{\mbox{$n_{e_o}$}}

\begin{document}

\title{Cosmological constraints on galaxy cluster structure}

\author{R. F. L. Holanda$^{1,2}$\footnote{E-mail: holanda@uepb.edu.br}}

\author{J. S. Alcaniz $^{3}$\footnote{E-mail: alcaniz@on.br}}

\address{$^1$Departamento de F\'{\i}sica, Universidade Estadual da Para\'{\i}ba, 58429-500, Campina Grande - PB, Brasil}

\address{$^2$Departamento de F\'{\i}sica, Universidade Federal da Para\'{\i}ba, Jo\~{a}o Pessoa - PB, Brasil}

\address{$^3$Departamento de Astronomia, Observat\'orio Nacional, 20921-400, Rio de Janeiro - RJ, Brasil}

\date{\today}

\begin{abstract}

Observations of galaxy clusters (GC's) are a powerful tool to probe the evolution of the Universe at $z<2$. However, the real shape and structure of these objects are not completely understood and the assumption of asphericity is often used in several cosmological tests. In this work, we propose methods to infer the elongation of the gas distribution of GC's based on measurements of the cosmic expansion rate, luminosity distance to type Ia supernovae, X-Ray and Sunyaev-Zeldovich properties of GC's and on the validity of the so-called distance duality relation. For the sample considered, we find that the clusters look preferentially elongated along the line of sight with the results of the different methods showing a good agreement with each other and with those predicted by the current cosmic concordance model.

\end{abstract}

\maketitle
%
%
%

\section{Introduction}\label{sec:introduction}

Galaxy clusters are the largest virialized astronomical structures in the Universe  and observations of their physical properties can provide important cosmological information. For instance, the matter density, $\Omega_m$, and the normalization of the density fluctuation power spectrum, $\sigma_8$, can be constrained from the evolution of galaxy clusters X-ray temperatures and their X-ray luminosity function (Henry 2000; Ikebe et al. 2002; Mantz et al. 2008; Vanderline et al. 2010). The abundance of galaxy clusters as a function of mass and redshift is expected to impose limits on evolution of $\omega$, the dark energy equation-of-state parameter, with statistical errors competitive with other techniques (Albrecht et al. 2006, Basilakos, Plionis \& Lima 2010; Chandrachani Devi \& Alcaniz, 2014). Assuming that the gas mass fraction does not evolve with redshift, X-ray observations of galaxy clusters can also be used as standard rulers to constrain cosmological parameters (Sasaki  1996; Pen 1997; Ettori et al.
 2003; Allen et al. 
2008; Gon\c calves et al. 2012) whereas the combination of the X-ray emission of the intracluster medium  with the Sunyaev-Zeldovich  effect (SZE) provide estimates of the angular diameter distance to the cluster redshift (Reese et al. 2002, De Filippis et al. 2005; Bonamente et al. 2006). Moreover, multiple redshift image systems behind  galaxy clusters  can also be used to estimate cosmological parameters via strong gravitational lensing (Lubini et al. 2013).

The intrinsic three-dimensional  galaxy clusters shape  is an important astrophysical probe and several methods to access this information have been proposed and tested with simulated data sets (Zaroubi et al. 1998; Reblinsky 2000; Ameglio et al. 2007; Allison et al. 2011). Particularly, Samsing et al. (2012) proposed a method  to measure the three-dimensional shape of the intracluster via X-ray observations only. This method, however, requires data of impressively high quality which is incompatible with current instruments. From the observational side, the first determination of the intrinsic triaxial shapes and tree-dimensional physical parameters of both dark matter (DM) and intra-cluster medium  was presented by Morandi et al. (2010) by combining X-ray, weak-lensing and strong-lensing observations. The Abell 1689 cluster can be modeled as a triaxial galaxy cluster with DM halo axial ratios $1.24 \pm 0.13$ and $2.37 \pm 0.11$ on the plane of the sky and along the line of sight, respectively. A model-
independent expression for the minimum line-of-sight extent of the hot plasma in a cluster of galaxies was proposed by Mahdavi \& Cheng (2011)  without assumptions regarding equilibrium or geometry. The only inputs were the 1-5 keV X-ray surface brightness and the Comptonization from SZE data. The method has been applied  to the Bullet Cluster implying minimum line-of-sight to plane of the sky axial ratio of $\approx 1$ (see Limousin et al . 2013 for an excellent review about galaxy cluster structure). From the cosmological side,  the so-called  distance-duality (DD) relation (Etherington 1933, Ellis 2007), i.e., $D_L(1+z)^{-2}/D_A=1$, where $D_L$ and $D_A$ are, respectively, luminosity and angular diameter distances to a given redshift $z$, has also been used to test different morphological models of galaxy clusters. In this case,  the 2-D ellipsoidal model was found to be the best geometrical hypothesis if the validity of the DD relation is assumed in cosmological observations (Holanda, Lima \& Ribeiro 
2010, 2011, 2012; Nair, Jhin-gan, \& Jain 2011; Meng, Zhang, Zhan \& Wang 2012). However, no information about cluster structure and shape was obtained so far using the properties of the DD relation.

In a series of papers, De Filippis and co-workers (2005) discussed a method to constrain the intrinsic shapes of galaxy clusters by combining their X-Ray and Sunyaev-Zeldovich effect (SZE) observations. By using  an isothermal triaxial $\beta$-model to gas distribution, De Filippis et al. (2005) argued that if the galaxy cluster is aligned along the line-of-sight (l.o.s.) and presents deviations from spherical symmetry, the angular diameter distance obtained via SZE/X-ray technique ($D_{A|Exp}$) furnishes $D_{A|Exp}=D_{A|true}e_{l.o.s.}$, where $e_{l.o.s}$ stands for the ratio of the radius of the cluster along the l.o.s. to its major axis perpendicular to the line of sight (spherical clusters have $e_{l.o.s}=1$). This method was originally applied to 25 galaxy clusters.  By assuming  $D_{A|true}$ as the one provided by the $\Lambda$CDM model they found that a large majority of the clusters  exhibit a marked triaxial structure.

Our goal in this paper is twofold. First, to propose a new approach based on the De Filippis et al. (2005) method to infer $e_{l.o.s.}$ of the gas distribution of galaxy clusters by using $D_{A|true}$ obtained directly from 26 independent measurements of the Hubble parameter between redshifts $0.07 < z < 2.3$.  Second, to show how  the assumption of the validity of the cosmic DD relation can be used to obtain $e_{l.o.s.}$ for each galaxy cluster via luminosity distances from type Ia supernovae (SNe Ia) observations. Finally, we update the De Filippis et al. (2005) results by using the most recent cosmic concordance  model based on the Plank data. In our analysis, we use a sample of 25 galaxy clusters and find an excellent agreement between the estimates of $e_{l.o.s.}$ from these different approaches.

\begin{figure*}[t]
\label{Fig1}
\centerline{
\psfig{figure=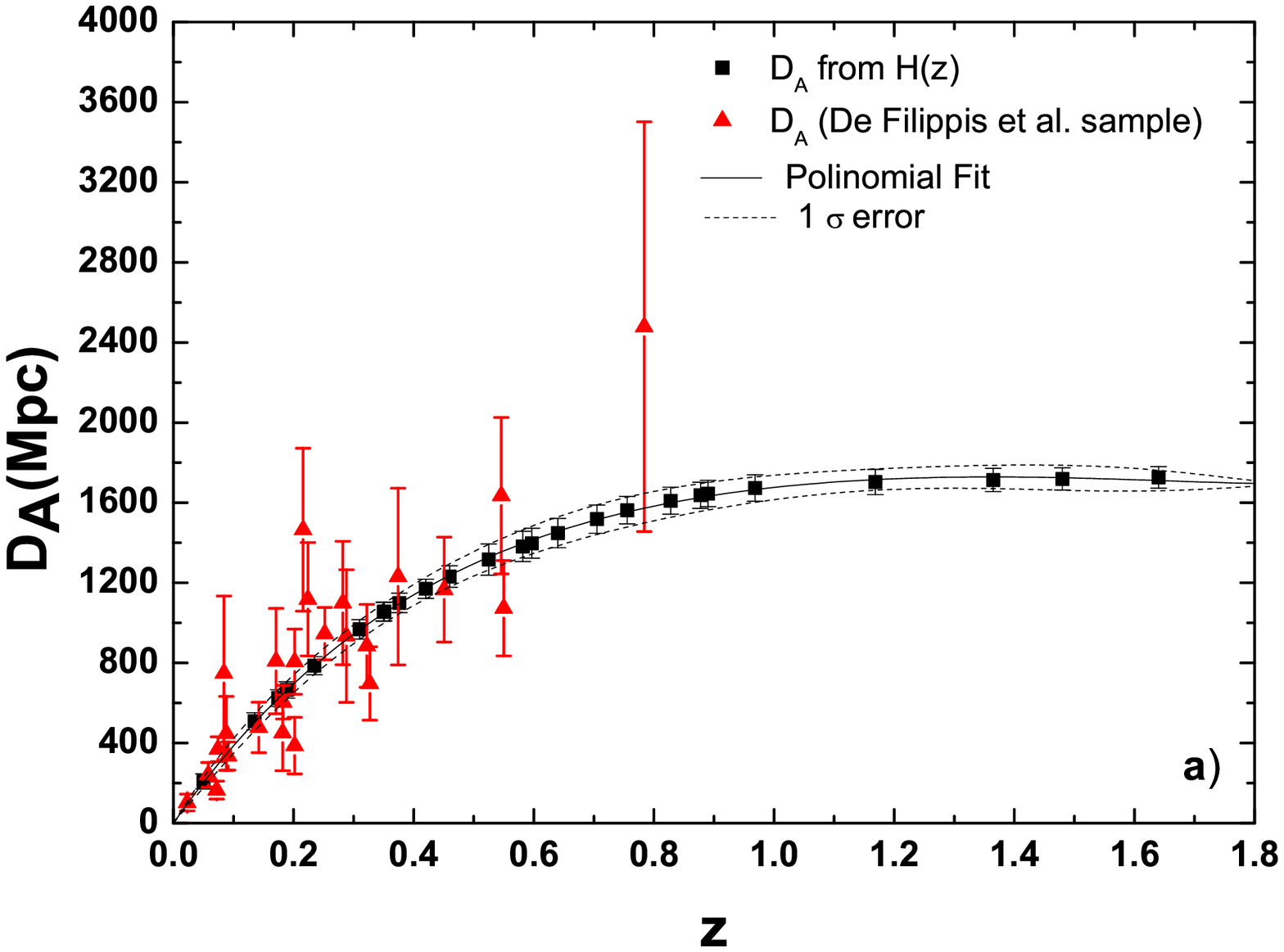,width=0.5\textwidth}
\psfig{figure=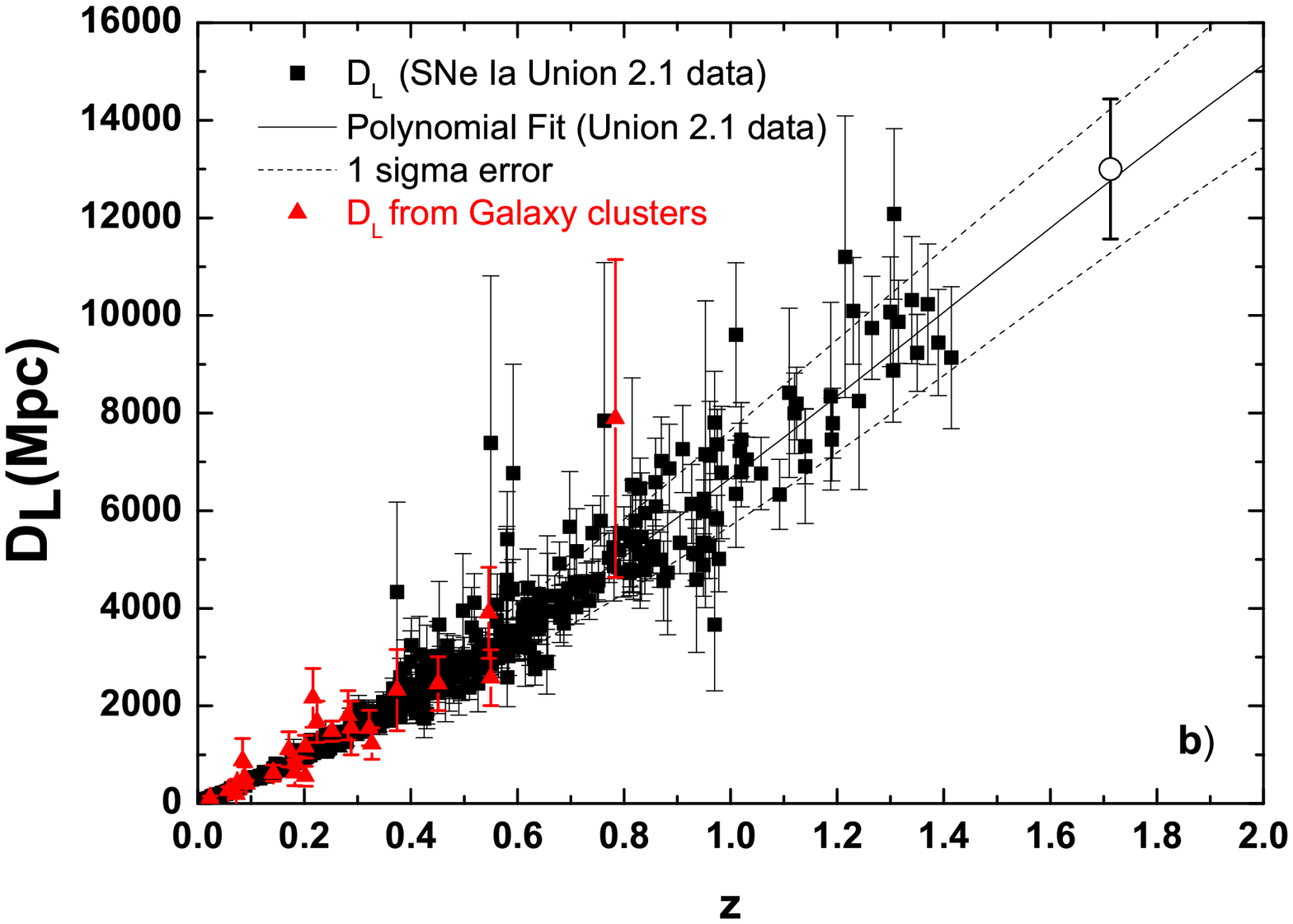,width=0.5\textwidth}
\hskip 0.1in}
\caption{{{a)}} Angular diameter distance from Galaxy clusters (red triangles) and from $H(z)$ measurements in a flat Universe (black squares). {{b)}} Luminosity distance of SNe Ia (Union 2.1 compilation from Susuki et al. 2012) (black squares) and from Galaxy clusters (red triangles). The open circle is the  SCP-$0401$ SNe Ia. In both figures the solid and dashed lines are, respectively, the best fit and 1 $\sigma$ error of the polynomial fit. }

\end{figure*}

\section{Cluster elongation along the line of sight}

\subsection{Method I}

We call method I the one based on De Filippis et al. (2005) approach. As commented earlier, these authors proposed a method to constrain the intrinsic shapes of galaxy clusters by combining X-ray and Sunyaev-Zeldovich observations. As it is well known, when the X-ray surface brightness is combined with the SZE temperature decrement in the cosmic microwave background spectrum, the angular diameter distance of galaxy clusters can be obtained. Thus, by using a triaxial elliptical $\beta$ model, De Filippis et al. (2005)  obtained a more general expression of the angular diameter distance, i.e.,
\begin{eqnarray}
D_{A|True} & =&  \left. D_{\rm A}\right|_{\rm Exp}
\frac{\theta_{\rm c,proj}}{\theta_{\rm c3}}h^{1/2} \nonumber \\ &  = & \left. D_{\rm
A}\right|_{\rm Exp} h^{3/4} \left(\frac{e_{\rm proj} }{e_1 e_2}
\right)^{1/2}\;,
\label{eq:dis2}
\end{eqnarray}
where $e_{proj}$ is the ratio of the major to the minor axes of the observed projected isophotes, $e_1 $ and $e_2 $ are axial ratios with respect to axis along l.o.s for each cluster and $h$ is a function of the cluster shape and orientation. $\left. D_{\rm A}\right|_{\rm Exp}$ is an observational quantity given by:
\begin{eqnarray}
\label{eq:obl7}
\left. D_{\rm A}\right|_{\rm Exp} &= &\frac{\Delta T_0^2}{S_{\rm X0}}
\left( \frac{m_{\rm e} c^2}{k_{\rm B} T_{e0} } \right)^2 \frac{g\left(\beta\right)}{g(\beta/2)^2\ \theta_{\rm c,proj}}\nonumber \\
& \times &\frac{\Lambda_{eH0}\ \mu_e/\mu_H}{4 \pi^{3/2}f(\nu,T_{\rm e})^2\ T^2_{\rm CMB}\ \sigma_{\rm T}^2\ (1+z_{\rm c})^4},
\end{eqnarray}
where $\Delta T_0$ is the central temperature decrement, $T_{\rm e}$ is the temperature of the ICM, $k_{\rm B}$ is the Boltzmann constant, $T_{\rm CMB} =2.728^{\circ}$K is the present-day temperature of the CMB, $\sigma_{\rm T}$ is the Thompson cross section, $m_{\rm e}$ is the electron mass, $c$ is the speed of light in vacuum, $\Lambda_{eH}$ is the X-Ray cooling function of the ICM in the cluster rest frame,  $S_{X0}$ is the  central surface brightness, $\mu$ is the molecular weight, $g(\beta)\equiv\frac{\Gamma \left[3\beta-1/2\right]}{\Gamma \left[3 \beta\right]}$ ($\Gamma$ denotes the Gamma function) and $f(\nu, T_{\rm e})$ accounts for frequency shift and relativistic corrections in SZE. The quantities $h$, $e_1$ and $e_2$ are not directly accessible. However, by considering that the cluster is aligned along the l.o.s.\footnote{As emphasized by De Filippis et al. (2005), the error due to inclination issues on the estimate of axial ratios can be usually neglected with respect to other uncertainties.}  we have $h=1$ and $D_{A|true} = D_{A|Exp}/e_{l.o.s.}$, where $e_{l.o.s.}$ is the the elongation, defined as the ratio of the radius of the cluster along the l.o.s. to its major axis along the perpendicular of sight (p.o.s.). Clusters which are instead more or less elongated along the l.o.s. than along  the p.o.s. will have values of $e_{\rm l.o.s.} > 1$ or $e_{\rm l.o.s.} < 1$, respectively.

In our analysis, $D_{A|true}$ is not calculated from a given cosmological model as in De Filippis et al. (2005). Here, we follow the method firstly presented by Holanda, Carvalho \& Alcaniz (2013) in which  $H(z)$ measurements are transformed into angular distance estimates by solving numerically the comoving distance integral for non-uniformly spaced data, i.e.,
\begin{equation}
\label{eq2}
D_C =c \int_0^z{dz^\prime \over H(z^\prime)}\approx {c\over 2}\sum_{i=1}^{N} (z_{i+1}-z_i)\left[ {1\over H(z_{i+1)}}+{1\over H(z_i)} \right].
\end{equation}
As one may check, by using standard error propagation techniques, the error associated to the $i^{th}$ bin is given by
\begin{equation}
s_i={c\over 2}(z_{i+1}-z_i)\left({\sigma_{H_{i+1}}^2\over H_{i+1}^4} + {\sigma_{H_{{i}}}^2\over H_{i}^4}\right)^{1/2}\;,
\end{equation}
where we have only taken into account the uncertainty on the values of $H(z)$ since the error on $z$ measurements is negligible. Therefore, the error of the integral (\ref{eq2}) in the interval $z=0$ -- $z_{n}$ is $\sigma^2_n = \sum_{i=1}^n s_i$.  We use 26 independent $H(z)$ measurements (Gaztanaga, Cabré \& Hui 2009; Stern et al. 2010; Moresco et al. 2012) in the redshift range $ 0 < z < 2.3$. For the current expansion rate, $H_0$, we use in our analysis $68 \pm 2.8$ km/s/Mpc, as derived from a median statistics analysis of 553 measurements of $H_0$ (Chen \& Ratra 2011), and which is in agreement with the value recently derived by the Planck Collaboration (Ade et al. 2013). As motivated by the WMAP and Planck results, we restrict our analysis to a spatially flat universe, such as $D_{A|True}$ is $D_{A|True}=\frac{1}{(1+z)}D_C$. 
\begin{figure*}[t]
\label{Fig1}
\centerline{
\psfig{figure=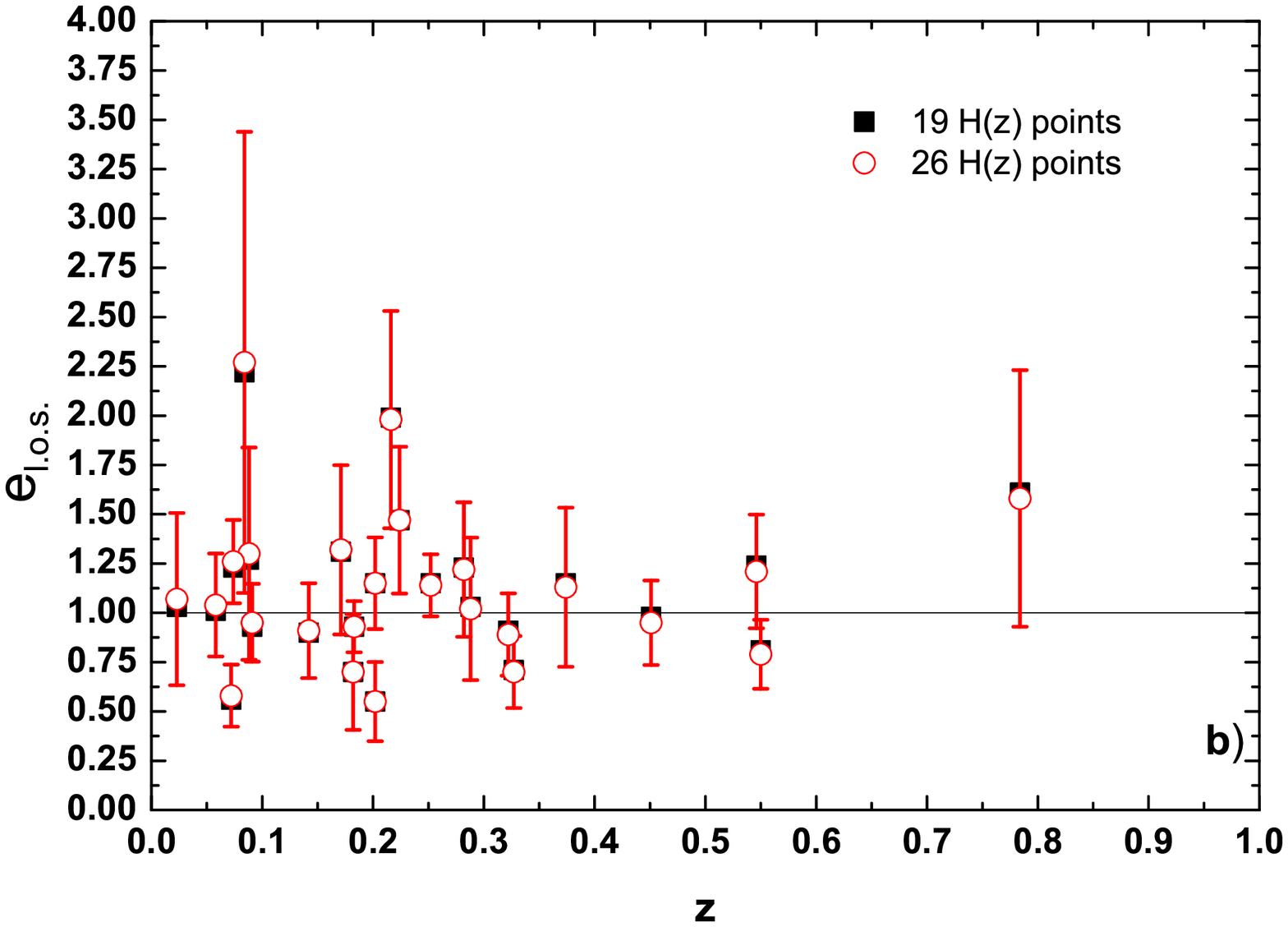,width=0.5\textwidth}
\psfig{figure=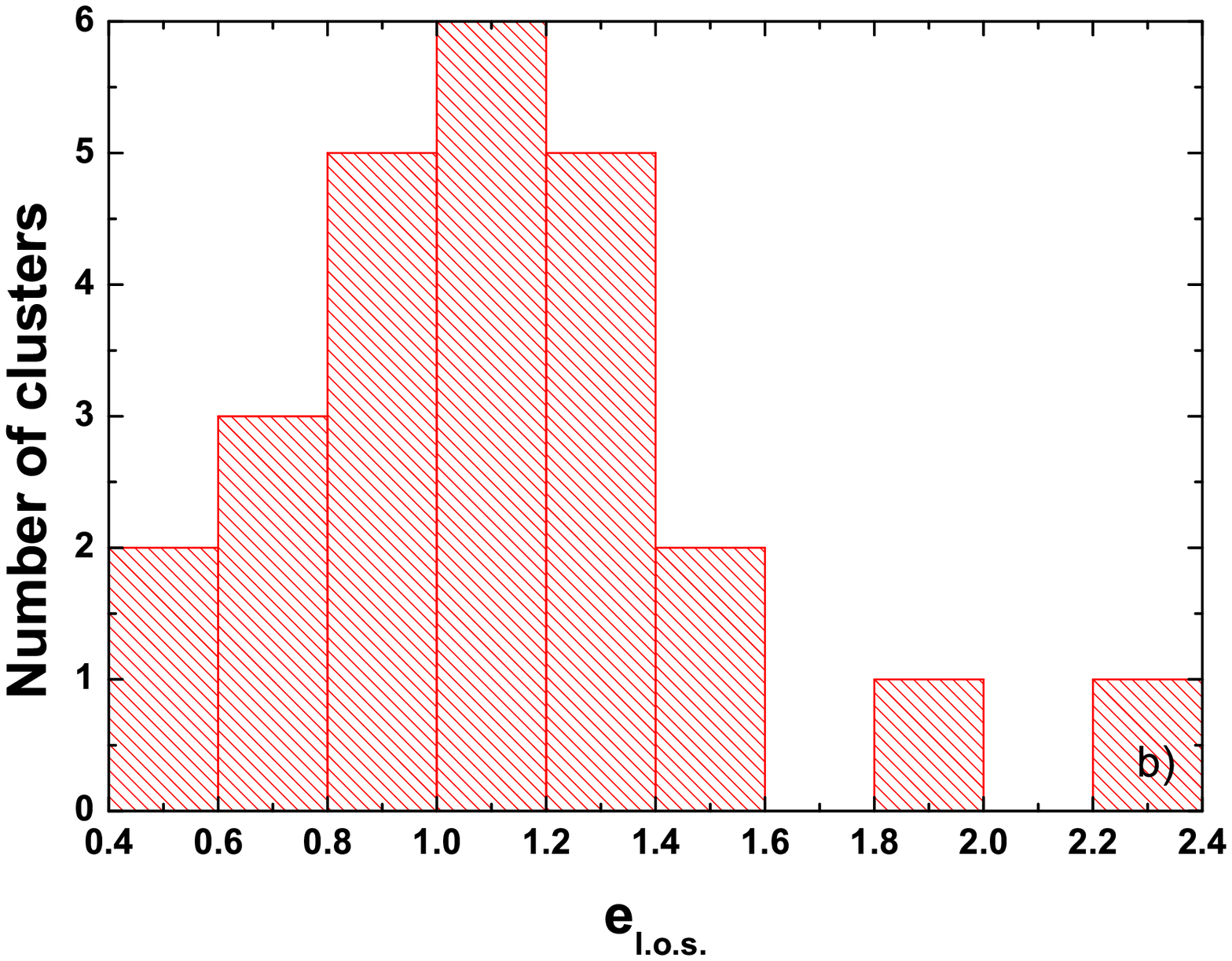,width=0.5\textwidth}
\hskip 0.1in}
\caption{{{a)}} $e_{l.o.s.}$ obtained from galaxy clusters + $H(z)$ analysis. {{b)}} Distribution of the galaxy cluster elongation along the l.o.s. for all clusters
in De Filippis et al. (2005) sample. }
\end{figure*}

In Fig.(1a) we show the galaxy cluster data of De Filippis et al. (2005) sample and $D_{A|True}$. The galaxy cluster sample comprises 25 galaxy clusters in the redshift range $0.023 \leq z \leq 0.8$ {for which archival X-ray data from XMM-Newton and Chandra observations were reanalyzed}\footnote{The first X-ray and SZE analyses were performed by Reese et al.\ (2002) and Mason et
al.\ (2001) where a spherical $\beta$ model was used to describe the clusters.} For this purpose, De Filippis et al. (2005) used an
isothermal elliptical $\beta$-model  to obtain $D_A(z)$ measurements for these galaxy clusters. { Observationally, the choice of circular rather than elliptical $\beta$ model does not affect the resulting of the Sunyaev-Zeldovich decrement ($\Delta T_0$) or central surface brightness ($\Delta T_0$) and the slope $\beta$ differs slightly between these models. However, different values for core radius ($r_c$) and, consequently, different $D_A(z)$ to galaxy clusters are obtained with these two galaxy cluster models (see Fig. 1 in their paper)}. Since the galaxy clusters and $H(z)$ observations are performed at different $z$, we calculate  $D_{A|True}$ at each galaxy cluster  redshift from a  polynomial fit of the points shown in Fig.(1a). 

In order to obtain $e_{l.o.s.}$ for each cluster, we use a maximum likelihood determined by a $\chi^2$ statistics,
\begin{equation}
\chi^2 =  \frac{(e_{l.o.s.} - \frac{D_{A|Exp}}{D_{A|true}})^2}{\sigma^2},
\end{equation}
where $\sigma^2$ are the errors associated with the observational techniques, i.e., \footnote{The error  is obtained by using standard error propagation for the quantity $D_{{A|Exp}}/{D_{A|true}}$.}

\begin{eqnarray}
\sigma^2 &=& \left[\frac{D_{A|Exp}}
{D^2_{A|true}}\right]^2 \sigma^2_{D_{A|true}}+ \left[\frac{1}{D_{A|true}}\right]^2\sigma^2_{D_{A|Exp}}.
\end{eqnarray}
For the galaxy cluster sample, the statistical error contributions are: SZE point sources $\pm 8\%$, X-ray background $\pm2\%$, Galactic NH $< \pm1\%$,
$\pm15\%$ for cluster asphericity, $\pm8\%$ kinetic SZ and for CMB anisotropy  $<\pm2\%$. Estimates for systematic effects are as follow: SZ calibration $\pm8\%$, X-ray flux calibration $\pm5\%$, radio
halos $+3\%$ and X-ray temperatute calibration $\pm7.5\%$. We also emphasize that typical statistical errors amount for nearly $20\%$ in agreement with other works (Mason et al. 2001; Reese et al. 2002, Reese 2004), while for systematic we also find typical errors around $+12.4\%$ and $-12\%$ (for more details, see Bonamente et al. 2006).

 \subsection{Method II}

The method II is directly linked with the cosmic DD relation, $D_L(1+z)^{-2}/D_A=1$. At this point, however, it is worth mentioning that two approaches have been adopted in the literature concerning the validity of this relation. Although a number of analysis have recently devoted themselves to establishing whether or not the DD relation holds in practice using current observational data (see, e.g., Holanda et al. 2013 and references therein), the majority of the studies in observational and theoretical cosmology assume this expression to be valid (see Shafieloo et al. 2013 for a discussion). Since the expected deviations from this relation are very small compared with the current observational uncertainties, we will adopt the second approach in our analysis to access the galaxy cluster elongation\footnote{The DD relation is independent either upon Einstein field equations or the nature of the matter content. Its validity requires only conservation of photon number and sources and observers to be 
connected by 
null geodesics in a general Riemannian spacetime. Searches for DD violation are, therefore, of great importance since any observational detection of it would be a clear evidence of new physics.}.
\begin{figure*}[t]
\label{Fig1}
\centerline{
\psfig{figure=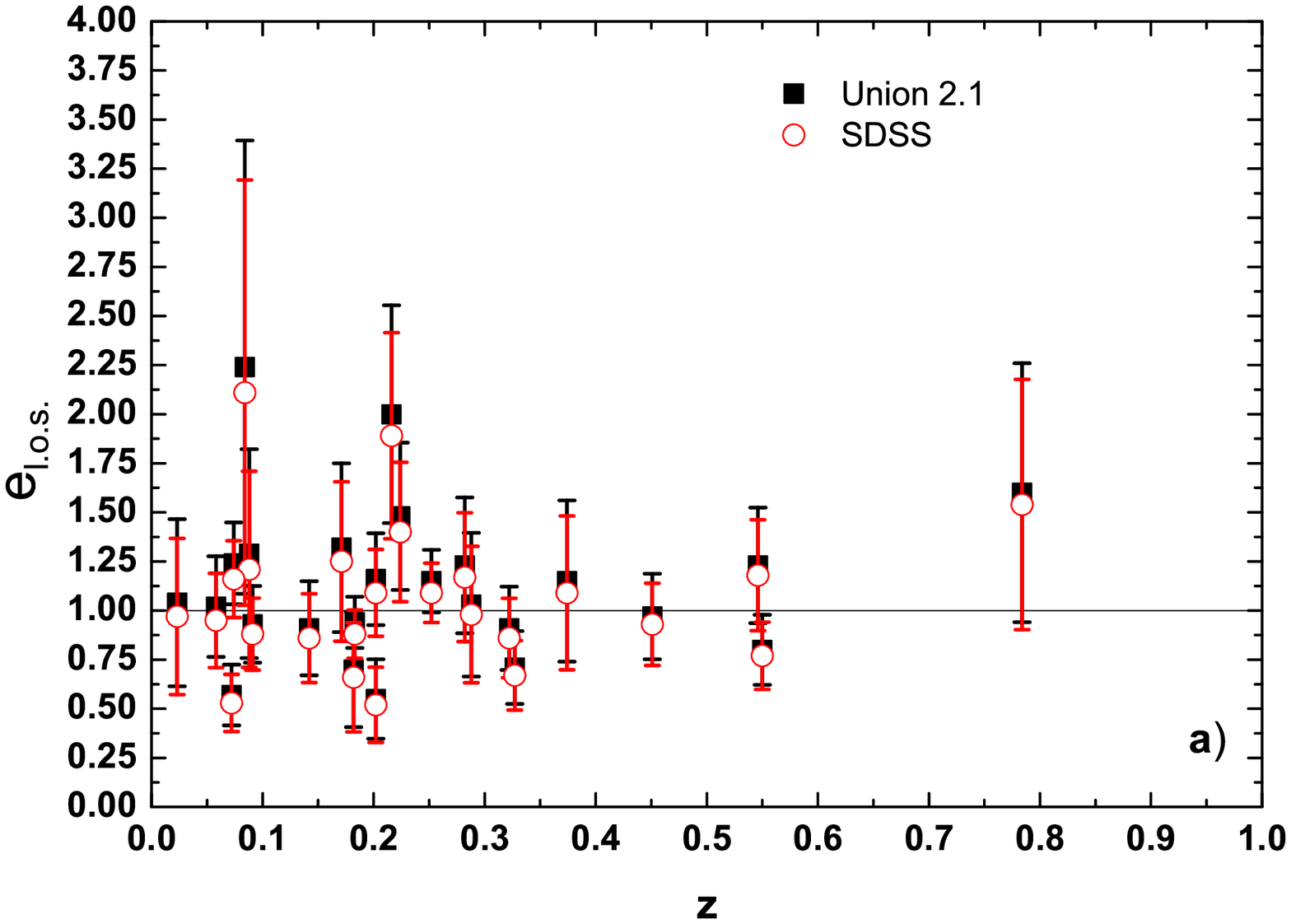,width=0.5\textwidth}
\psfig{figure=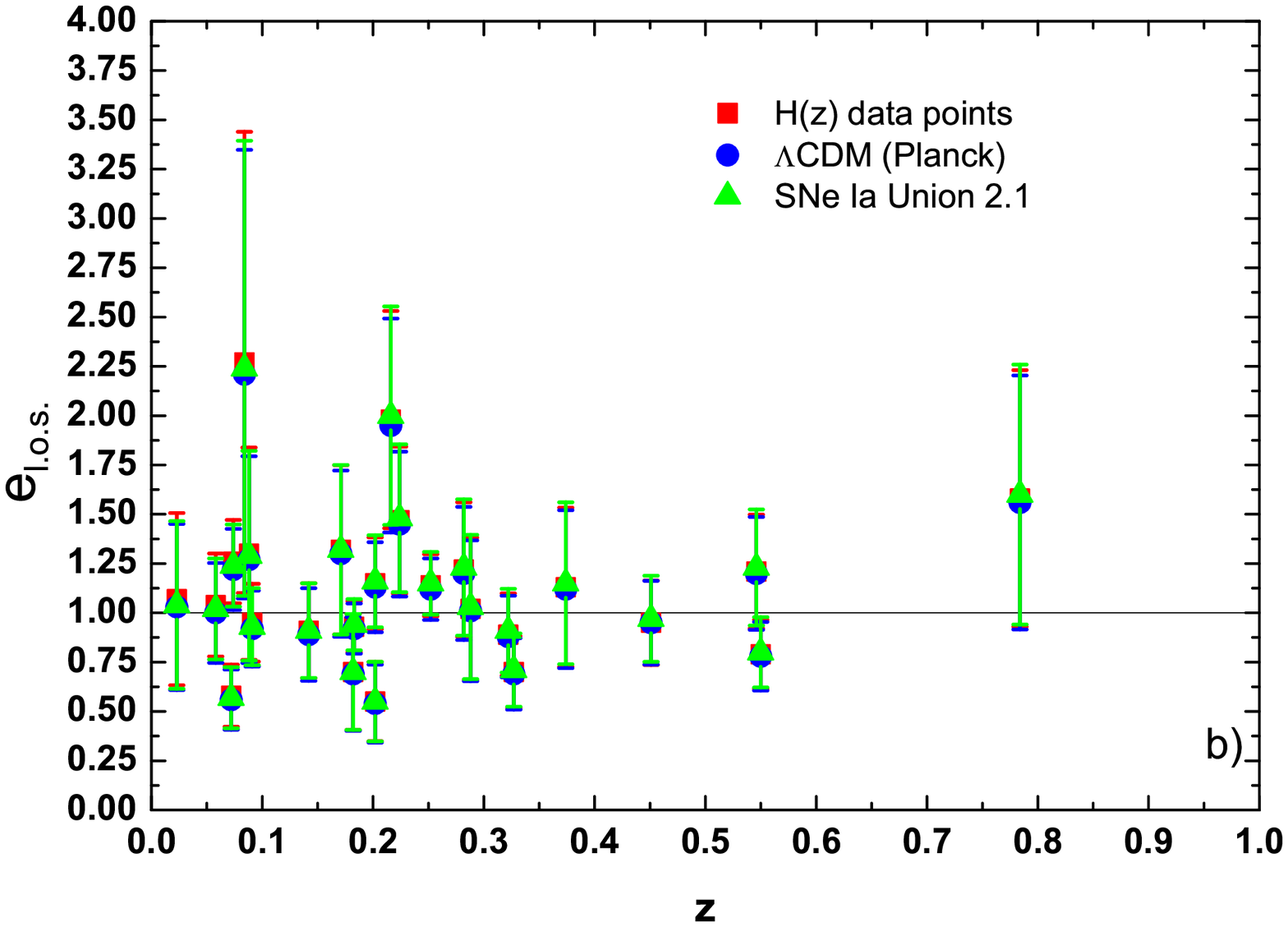,width=0.5\textwidth}
\hskip 0.1in}
\caption{ $e_{l.o.s.}$ obtained from galaxy clusters + SNe Ia analysis. {{b)}} Comparing  $e_{l.o.s.}$ obtained from galaxy clusters + $H(z)$, galaxy clusters + SNe Ia and galaxy clusters + $\Lambda$CDM prior. }
\end{figure*}

Estimates of the angular diameter distance from galaxy clusters are based on SZE/X-ray technique, such that $D_{A|Exp}=D_{A|true}e_{l.o.s.}$. Therefore, we can rewrite the DD relation to $D_L(1+z)^{-2}/D_{A|Exp}=1/e_{l.o.s}$ and, by using luminosity distances of SNe Ia, we can obtain the elongation for each galaxy cluster. In our analysis we use two SNe Ia compilations: the Union 2.1 (Suzuki et al. 2012), composed by 580 SNe Ia in the redshift range $0.015<z<1.43$, where we add the most distant ($z=1.713$) spectroscopically confirmed SNe Ia (SNe Ia SCP-$0401$, Rubin et al. 2013) and SDSS compilation (Kessler et al. 209), composed by 288 SNe Ia in the redshift range $0.02 \leq  z \leq 1.55$. These SNe Ia samples were calibrated with both SALT2 (Guy et al. 2007)  and MLCS2K2 (Jha et al.\ 2007) light-curve fitters. In this way, we explore the influence of different fits to the SNe Ia light-curves on our results. Again, since the galaxy clusters and SNe Ia observations are performed at different $z$, we 
calculate  $D_{L}$ at each galaxy cluster redshift from a  polynomial fit of the points shown in Fig.(1b).

In order to obtain $e_{l.o.s.}$ for each cluster, we use a $\chi^2$ statistics,
\begin{equation}
\chi^2 =  \frac{(e_{l.o.s.} - \frac{D_L}{(1+z)^{2}D_{A|Exp}})^2}{\sigma^2},
\end{equation}
where $\sigma^2$ are the errors associated with the observational techniques, i.e.,
\begin{eqnarray}
\sigma^2 &=& \left[\frac{D_L}
{(1+z)^{4}D^2_{A|Exp}}\right]^2 \sigma^2_{(1+z)^{2}D_{A|Exp}} \\ & &+ \left[\frac{1}{(1+z)^{2}D_{A|Exp}}\right]^2\sigma^2_{D_L} \nonumber.
\end{eqnarray}

\section{Results }

The results of our analysis are displayed in Figs. (2) and (3). In Fig.(2a) we plot the results obtained from the method I. The $H(z)$ sample comprises 19 measurements from the cosmic chronometers method (Simon et al. 2005, Stern et al. 2010, Moresco et al. 2012) and 7 $H(z)$ measurements from observations of baryon acoustic oscillations (BAOs) (Blake et al. 2012, Reid et al. 2012, Xu et al. 2013, Busca et al. 2013, Chuang \& Wang 2013). Black squares and open red circle correspond, respectively, to the complete (26 $H(z)$ data points) and the reduced (19 $H(z)$ points) samples, respectively. Clearly, both data sets provide very similar constraints for $e_{l.o.s.}$. In Fig.(2b) we plot the histogram of the distribution of elongation along the l.o.s. for all clusters. The clusters present elongated preferentially along the l.o.s. with mean $1.12\pm 0.08$. 

In Fig.(3a) we plot the results from method II. Black squares and open red circle correspond to the results obtained from the Union 2.1 (added with the SNe Ia SCP-$0401$, SALT2) and SDSS (MLCS2K2) SNe Ia samples, respectively. We see that both SNe Ia data sets provide very similar results, which means  that SNE Ia light-curve fitters seems not to influence the results. We find that the mean of the distributions of the elongation is $1.12\pm 0.08$ and $1.06\pm 0.07$ for SALT2 and MLCS2K2, respectively. 

 { In order to compare our results with those obtained in previous analysis, we update the De Filippis et al. (2005) results using the best-fit $\Lambda$CDM model from Planck collaboration (flat universe with $\Omega_m=0.315 \pm 0.017$ and $H_0 = 67.3 \pm 1.2$ km/s/Mpc) as a prior to evaluate $D_{A|True} $. For this case, we find that the mean of the distribution of the elongation is $1.10\pm 0.08$. In Fig.(3b) we plot the $e_{l.o.s.}$ results from methods I and II along with the result obtained assuming the best-fit $\Lambda$CDM model. An excellent concordance among these results is found. The fact that 15 clusters out of the 25 considered are more elongated than the 10 remaining is also  in perfect agreement with the De Filippis et al. (2005) results. These results are summarized in Tables I and II.}

\begin{table}
\begin{center}
\caption{Values of the distributions of GC's elongation (1$\sigma$)}
 {\begin{tabular} {@{}cccc@{}} Galaxy Cluster   & Method 1  & Method 2  & $\Lambda$CDM \\ 
                                                &  (26 H(z) data) & (Union 2.1) & Planck Results \\  \hline
																								\\ 

MS 1137.5+6625   &$1.57\pm{0.65}$  &$1.60\pm{0.66}$  &$1.56\pm{0.64}$  \\ 
MS 0451.6-0305   &$0.79\pm{0.17}$& $0.80\pm{0.18}$&$0.78\pm{0.17}$ \\ 
Cl 0016+1609	 &$1.20\pm{0.28}$& $1.22\pm{0.30}$ &$1.20\pm{0.28}$\\
RXJ1347.5-1145   &$0.95\pm{0.21}$  &$0.97\pm{0.28}$&$0.95\pm{0.21}$ \\
A 370            & $1.12\pm{0.40}$ & $1.14\pm{0.41}$&$1.12\pm{0.28}$ \\ 
MS 1358.4+6245	 & $0.69\pm{0.18}$ & $0.71\pm{0.18}$&$0.69\pm{0.18}$ \\ 
A 1995		 & $0.89\pm{0.20}$ & $0.91\pm{0.21}$&$0.88\pm{0.20}$ \\ 
A 611		 & $1.01\pm{0.36}$ & $1.03\pm{0.36}$ &$1.00\pm{0.35}$\\ 
A 697		 & $1.22\pm{0.34}$ & $1.23\pm{0.34}$ &$1.20\pm{0.33}$\\ 
A 1835		 & $1.13\pm{0.16}$ & $1.14\pm{0.16}$&$1.12\pm{0.16}$ \\ 
A 2261		 & $1.47\pm{0.37}$ & $1.48\pm{0.37}$ &$1.45\pm{0.36}$\\ 
A 773		 & $1.98\pm{0.55}$ & $2.00\pm{0.55}$ &$1.95\pm{0.54}$\\ 
A 2163		 & $1.15\pm{0.23}$ & $1.15\pm{0.23}$&$1.12\pm{0.22}$ \\ 
A 520		 & $0.55\pm{0.20}$ & $0.55\pm{0.20}$ &$0.54\pm{0.20}$\\ 
A 1689		 & $0.93\pm{0.13}$ & $0.94\pm{0.13}$&$0.92\pm{0.13}$ \\ 
A 665		 & $0.70\pm{0.30}$ & $0.70\pm{0.29}$ &$0.69\pm{0.28}$\\ 
A 2218		 & $1.30\pm{0.54}$ & $1.31\pm{0.43}$&$1.29\pm{0.42}$ \\ 
A 1413		 & $0.91\pm{0.24}$ & $0.90\pm{0.23}$&$0.89\pm{0.23}$ \\ 
A 2142		 & $0.94\pm{0.20}$ & $0.93\pm{0.19}$&$0.92\pm{0.19}$ \\ 
A 478		 & $1.30\pm{0.54}$ & $1.28\pm{0.53}$ &$1.26\pm{0.52}$\\ 
A 1651		 & $2.27\pm{1.16}$ & $2.23\pm{1.15}$&$2.20\pm{0.13}$\\ 
A 401		 & $1.25\pm{0.21}$ & $1.23\pm{0.20}$ &$1.21\pm{0.20}$\\ 
A 399		 & $0.58\pm{0.16}$ & $0.57\pm{0.15}$ &$0.56\pm{0.15}$\\ 
A 2256		 & $1.03\pm{0.26}$ & $1.01\pm{0.25}$ &$1.00\pm{0.25}$\\ 
A 1656		 & $1.07\pm{0.43}$ & $1.04\pm{0.42}$ &$1.02\pm{0.42}$\\
\hline
\end{tabular} \label{ta1}}
\end{center}
\end{table}

\begin{table}
\begin{center}
\caption{Mean values of the distributions of GC's elongation}
 {\begin{tabular} {@{}cc@{}}  Method   & $e_{l.o.s.}$ ($1\sigma$) \\ \hline \hline
I - H(z) &       $1.12\pm 0.08$ \\
I - $\Lambda$CDM         & $1.10\pm 0.08$\\
II - SNe Ia (Union 2.1)     &$1.12\pm 0.08$ \\
II - SNe Ia (SDSS)     &$1.06\pm 0.07$  \\
De Filippis et al. 2005  &$1.15 \pm 0.08$ \\
\hline
\end{tabular} \label{ta1}}
\end{center}
\end{table}

 \section{Conclusions}

In order to use galaxy clusters as a cosmological probe it is needed to add some complementary assumptions about their morphology and numerous studies about their intracluster gas and dark matter distributions have assumed standard spherical geometry (Reiprich \& Boringer 2002; Bonamente et al. 2006; Shang, Haiman \& Verdi 2009). Basically, this is due to the quality of the current observations which do not allow constraints on triaxial models. 

In this work, assuming that galaxy clusters are aligned along the line of sight and using an isothermal triaxial $\beta$-model to the gas distribution, we have proposed two independent ways to infer the structure of a sample of 25 galaxy clusters whose angular diameter distances were obtained from  X-ray and Sunyaev-Zeldovich observations, $D_{A|Exp}$. In method I, the elongation along the l.o.s., $e_{l.o.s.}$ for all clusters has been calculated comparing  $D_{A|True}$ obtained from 26 independent measurements of the Hubble parameter with $D_{A|Exp}$ of the galaxy clusters. {In the method II,  we have considered the validity of the cosmic DD relation and shown that it can be rewritten as $D_L(1+z)^{-2}/D_{A|Exp}=1/e_{l.o.s}$.} Then, using luminosity distances from SNe Ia observations, we have inferred the elongation of each galaxy cluster in a given sample. As a general result, we have obtained that the clusters of our sample seem to be  preferentially  elongated
along the l.o.s. with the mean of the distributions of the GC's elongation in agreement between the two methods (see Tables I and II). 

Since galaxy clusters structure and their clustering properties depend strongly upon the  underlying cosmology, we have also estimated the elongation distribution assuming the best-fit $\Lambda$CDM model given by the Planck Collaboration. We have found that the results are completely equivalent to the ones obtained from methods I and II. Finally, we believe that the methods and results presented here reinforce the idea of non-sphericity of galaxy clusters. Moreover, when larger samples with smaller statistical and systematic uncertainties of $H(z)$ measurements, SNe Ia as well as X-ray and SZE observations  become available it will be possible to improve the determination  of the intrinsic shape of galaxy clusters.

\section*{Acknowledgments}

R.F.L. Holanda is supported by INCT-A and CNPq (No. 478524/2013-7). J.S. Alcaniz is supported by INEspa\c{c}o, CNPq (305857/2010-0 and 485669/2011-0) and FAPERJ (E-26/103.239/2011).

\end{document}